\documentclass[11pt,fleqn]{article}
\usepackage{psfig}
\title{Possible Dark States induced by a Surface Wave \\ along a Vacuum-Matter Boundary}
\author{Zotin K.-H. Chu} 
\date{4-601, Building C, Beijingcheng,
Baixingkangcheng,  Changsha Road, Urumqi 830013, China and P.O.
Box 39, Distribution Unit,  Xihong Road, Urumqi 830000, China}
\begin{document}
\maketitle
\begin{abstract}
Possible dark states could be induced after  derivations of the
entrainment of matter induced by a surface  wave propagating along
the flexible vacuum-matter boundary  by considering the nonlinear
coupling between the interface and the rarefaction effect. The
nonrelativistic limit of the relativistic Navier-Stokes equations
was considered and analytically solved by a perturbation approach.
The critical reflux values associated with the product of the
second-order
 body forcing and the Reynolds number (representing the
viscous dissipations) decrease as the Knudsen number (representing
the rarefaction measure) increases from zero to 0.1. We obtained
the critical bounds for possible dark states corresponding to
specific Reynolds numbers (ratio of wave inertia and viscous
dissipation effects) and wave numbers which might be linked to the
dissipative evolution of certain large-scale structure during the
relativistic heavy-ion collisions.
\newline

\end{abstract}
\doublerulesep=6mm        
\baselineskip=6mm         %
\oddsidemargin-1mm
\bibliographystyle{plain}
\section{Introduction}
Recently a new state of matter has been created in Au+Au
collisions at RHIC, and surprisingly it was found to flow as a
perfect fluid [1]. The kinematic shear viscosity of this nearly
perfect fluid has been determined and found to be rather small
compared to conventional low-temperature fluids. Relevant
researches in heavy-ion physics has focused on constraining the
transport properties of hot and dense nuclear matter using
experimental data from RHIC and studying relativistic
hydrodynamics for viscous fluids in order to describe the
expansion of the fireballs created in relativistic heavy-ion
collisions.
\newline
Meanwhile the mean cosmic density of dark matter (plus baryons) is
now pinned down to be only ca. 30\% of the so-called critical
density corresponding to a 'flat'-Universe. However, other recent
evidence|microwave background anisotropies, complemented by data
on distant supernovae|reveals that our Universe actually is
'flat', but that its dominant ingredient (ca. 70\% of the total
mass energy) is something quite unexpected: 'dark energy'
pervading all space, with negative pressure.
We do know that this material is very dark and that it dominates
the internal kinematics, clustering properties and motions of
galactic systems. Dark matter is commonly associated to weakly
interacting particles (WIMPs), and can be described as a fluid
with vanishing pressure. It plays a crucial role in the formation
and evolution of structure in the universe and it is unlikely that
galaxies could have formed without its presence [2].
Analysis of cosmological mixed dark matter models in spatially flat
Friedmann
Universe with zero $\Lambda$ term have been presented before.
A large majority of dark energy models describes dark energy in
terms of the equation of state (EOS) $p_d = \omega\,\rho_d$ (cf.
Refs. 3 and 4), where $\omega$ is the parameter of the EOS, while
$p_d$ and $\rho_d$ denote the pressure and the energy density of
dark energy, respectively. The value $\omega = -1$ is
characteristic of the cosmological constant, while the dynamical
models of dark energy generally have $\omega \ge -1$. The case of
the growing cosmological term $\Lambda$ and its implications for
the asymptotic expansion of the universe and the destiny of the
bound systems have been studied in Ref. 4 using above system of
equations. Their results showed that even for very slow growth of
$\Lambda$ (which satisfies all the conditions on the variation of
$G_N$), in the distant future the gravitationally bound systems
become unbound, while the non-gravitationally bound systems remain
bound.
\newline
%
%
%
Influential only over the largest of scales-the cosmological
horizon-is the outermost species of invisible matter: the vacuum
energy (also known by such names as dark energy, quintessence,
$x$-matter, the zero-point field, and the cosmological constant
$\Lambda$) (cf. Refs. 5 and 6). If there is no exchange of energy
between vacuum and matter components, the requirement of general
covariance implies the time dependence of the gravitational
constant $G$. Thus, it is interesting to look at the interacting
behavior between the vacuum (energy) and the matter from the
macroscopic point of view. One related issue, say, is about the
dissipative matter of the flat Universe immersed in vacua [7] and
the other one is the macroscopic Casimir effect with the deformed
boundaries [8].
\newline
%
Theoretical (using the Boltzmann equation) and experimental
studies of interphase nonlocal transport phenomena which appear as
a result of a different type of nonequilibrium representing
propagation of a surface elastic wave have been performed since
late 1980s (cf. Refs. 9 and 10). These are relevant to rarefied
gases (RG) flowing along deformable elastic slabs with the
dominated parameter being the Knudsen number (Kn =
mean-free-path/$L_d$, mean-free-path (mfp) is the mean free path
of the gas, $L_d$ is proportional to the distance between two
slabs) [11-13]. The role of the Knudsen number is similar to that
of the Navier slip parameter $N_s$ (cf. Ref. 14); here, $N_s = \mu
S/d$ is the dimensionless Navier slip parameter; S is a
proportionality constant as $u_s = S \tau$, $\tau$ : the shear
stress of the bulk velocity; $u_s$ : the dimensional slip
velocity; for a no-slip case, $S = 0$, but for a no-stress
condition. $S=\infty$, $\mu$ is the fluid viscosity, $d$ is one
half of the distance between upper and lower slabs).
\newline Note that, there
are some models, like the MIT bag model and its descendants, where
matter is in a bag, in which there is no vacuum  and the vacuum is
outside. In this particular model one might speak of a clear {\it
vacuum-matter boundary}. Here, borrowing the idea of  the MIT bag
model, the transport driven by the wavy elastic vacuum-matter
boundary will be presented. The flat-Universe is presumed and the
corresponding matter is immersed in vacua with the interface being
flat-plane like. We adopt the macroscopic or hydrodynamical
approach and simplify the original system of equations (related to
the momentum and mass transport) to one single higher-order
quasi-linear partial differential equation in
terms of the unknown stream function. 
We then introduce the perturbation technique so that we can solve
the related boundary value problem approximately. To consider the
originally quiescent gas for simplicity, due to the difficulty in
solving a fourth-order quasi-linear complex ordinary differential
equation (when the wavy boundary condition are imposed), we can
finally get an analytically perturbed solution and calculate those
physical quantities we have interests, like, time-averaged
transport or entrainment, perturbed velocity functions, critical
unit body forcing corresponding to the possible dark states. These
results might be closely linked to the vacuum-matter interactions
(say, macroscopic Casimir effects) and the evolution of the
Universe (as mentioned above : the critical density [2]). Our
results also show that for certain time-averaged evolution of the
matter (the maximum speed of the matter (gas) appears at the
center-line) there might be existence of negative-pressure states.
\section{Formulations}
The matter is presumed to be a fluid associated with a shear
viscosity but no bulk viscosity and no heat conduction here (the
geometrized units are adopted  so that $G=c=1$ and the Einstein's
field equations : $G_{\mu\nu}=8\pi T_{\mu\nu}$). The stress tensor
(for this fluid) is [15]
\begin{displaymath}
 T_{\mu\nu}=(\rho_0+\rho_0 e_0 +P)u_{\mu}u_{\nu}+P
 g_{\mu\nu}-2\eta \sigma_{\mu\nu}.
\end{displaymath}
Here, $\rho_0$, $e_0$, $P$, and $u_{\mu}$ are the rest-mass
density, specific internal energy, pressure, and the fluid
$4$-velocity, respectively. $\eta$ is the coefficient of viscosity
and is related to the kinematic viscosity $\nu$ by $\eta=\rho_0
\nu$. $\sigma_{\mu\nu}$ is the shear tensor (the detailed
expression could be traced in Ref. 15). In general a $\Gamma$-law
equation of state $P=(\Gamma-1)\rho_0 e_0$ could be presumed.
Thus, we have basic fluid variables
\begin{displaymath}
 \rho_*=\rho_0 \alpha u^0 e^{6\phi}, \hspace*{6mm}
 e_*=(\rho_0 e_0)^{1/\Gamma}\alpha u^0 e^{6\phi}, \hspace*{6mm}
 \tilde{S}_k=\rho_* h u_k,
\end{displaymath}
where $\phi$ is the the conformal exponent and $h=1+e_0+P/\rho_0$
is the specific enthalpy ($\alpha$ is the elapse or proper time
elapsed in moving between the neighbouring spatial hypersurfaces).
The conservation of stress-energy [15] gives $T^{\mu\nu}_{;\nu}=0$
and the law of baryon number conservation $\nabla_{\mu} (\rho_0
u^{\mu})=0$ gives the relativistic continuity, energy, and
Navier-Stokes equations
\begin{equation}
 \partial_t \rho_* +\partial_i (\rho_* v^i)=0,
\end{equation}
\begin{equation}
 \partial e_* +\partial_i (e_* v^i)=\frac{2}{\Gamma}\alpha
 e^{6\phi} \eta (\rho_0 e_0)^{(1-\Gamma)/\Gamma}
 \sigma^{\alpha\beta} \sigma_{\alpha\beta},
\end{equation}
\begin{equation}
 \partial_t \tilde{S}_k +\partial_i (\tilde{S}_k v^i)=-\alpha
 e^{6\phi} P_{,k} +2 (\alpha e^{6\phi}\eta \sigma^{\mu}_k)_{,\mu}
 +\alpha e^{6\phi} g^{\alpha\beta}_{,k} (\eta
 \sigma_{\alpha\beta}-
 \frac{1}{2}\rho_0 h u_{\alpha}u_{\beta}),
\end{equation}
where $v^i=u^i/u_0$ is the $3$-velocity. The quantity $u^0$ is
determined by the normalization condition $u^{\nu}u_{\nu}=1$,
which has
\begin{displaymath}
 w^2=\rho_*^2+e^{-4\phi} \tilde{\gamma}^{ij}\tilde{S}_i\tilde{S}_j
 [1+\frac{\Gamma {e_*}^{\Gamma}}{\rho_* (w
 e^{6\phi}/\rho_*)^{\Gamma-1}}]^{-2},
\end{displaymath}
with $w=\rho_* \alpha u^0$ ($\gamma_{ij}$ is the spatial or
$3$-metric). We remind the readers that the stress tensor
$T^{\mu\nu}$ generates the source terms in the field evolution
equations :
\begin{displaymath}
 \rho=h w e^{-6\phi}-P-\frac{2\eta}{\alpha^2}(\sigma_{tt}-
 2\sigma_{ti}\beta^i+\sigma_{ij}\beta^i\beta^j),
\end{displaymath}
\begin{displaymath}
 S_i=e^{-6\phi}\tilde{S}_i-\frac{2\eta}{\alpha}(\sigma_{ti}-
 \sigma_{ij}\beta^j),
\end{displaymath}
\begin{displaymath}
S_{ij}=\frac{e^{-6\phi}}{w h}\tilde{S}_i\tilde{S}_j +P
\gamma_{ij}-2\eta \sigma_{ij},
\end{displaymath}
where $\beta^i$ is the shift (for gauge conditions) or
displacement in spatial coordinates in moving between the
neighbouring spatial hypersurfaces.\newline Equations (1-3) are
too difficult to be solved analytically or even by a perturbation
approach. In this work we only consider the nonrelativistic limit
of above equations. Meanwhile the vacuum is presumed to be
incompressible (cf., e.g., Ref. 16, the  speed of sound
propagating in this 'vacuum' is formally very large (or infinite)
rather than zero as in the empty vacuum; this implies the
incompressible vacuum).
\newline
Note that the first theories of relativistic dissipative fluid
dynamics are due to Eckart [17] and to Landau and Lifshitz [18].
The difference in formal appearance stems from different choices
for the definition of the hydrodynamical four-velocity. These
conventional theories of dissipative fluid dynamics are based on
the assumption that the entropy four-current contains terms up to
linear order in dissipative quantities and hence they are referred
to as {\it first order} theories of dissipative fluids. The
resulting equations for the dissipative fluxes are linearly
related to the thermodynamic forces, and the resulting equations
of motion are parabolic in structure, from which we get the
Fourier-Navier-Stokes equations. They have the undesirable feature
that {\it causality} may not be satisfied. That is, they may
propagate viscous and thermal signals with speeds exceeding that
of light. Extended theories of dissipative fluids due to Grad
[19], M\"{u}ller [20], and Israel and Stewart [21] were introduced
to remedy some of these undesirable features. These causal
theories are based on the assumption that the entropy four-current
should include terms quadratic in the dissipative fluxes and hence
they are referred to as {\it second order} theories of dissipative
fluids. The resulting equations for the dissipative fluxes are
hyperbolic and they lead to causal propagation of signals [22].
A qualitative study of relativistic dissipative fluids for
applications to relativistic heavy ions collisions has been done
using these first order theories. The application of second order
theories to nuclear collisions has just begun [23].
\newline
The flat-plane boundaries of this matter-region  or the
vacuum-matter boundaries are rather flexible and presumed to be
elastic, on which are imposed traveling sinusoidal waves of small
amplitude $a$ (possibly due to vacuum fluctuations). The vertical
displacements of the upper and lower interfaces ($y=h$ and $-h$)
are thus presumed to be $\eta$ and $-\eta$, respectively, where
$\eta=a \cos [2\pi (x-ct)/\lambda$], $\lambda$ is the wave length,
and $c$ the wave speed. $x$ and $y$ are Cartesian coordinates,
with $x$ measured in the direction of wave propagation and $y$
measured in the direction normal to the mean position of the
vacuum-matter interfaces. The schematic plot of above features is
shown in Fig. 1.\newline
It would be convenient to simplify these equations by introducing
dimensionless variables. We have a characteristic velocity $c$ and
three characteristic lengths $a$, $\lambda$, and $h$. The
following variables based on $c$ and $h$ could thus be introduced
:
 $x'={x}/{h}$, $ y'={y}/{h}$, $u'={u}/{c}$, $v'={v}/{c}$,
 $\eta'={\eta}/{h}$, $\psi'={\psi}/({c\,h})$,
 $t'={c\,t}/{h}$,  $p'={p}/({\rho c^2})$,
where $\psi$ is the dimensional stream function, $u$ and $v$ are the velocities
along the $x$- and $y$-directions; $\rho$ is the density, $p$ (its gradient)
is related to the (unit) body
forcing. The primes could be dropped in the following. The amplitude
ratio $\epsilon$, the wave number $\alpha$, and the Reynolds
number (ratio of wave inertia and viscous dissipation effects) $Re$ are defined by
\begin{displaymath}
 \epsilon=\frac{a}{h}, \hspace*{4mm} \alpha=\frac{2 \pi h}{\lambda},
 \hspace*{4mm} Re =\frac{c\,h}{\nu}.
\end{displaymath}
We shall seek a solution in the form of a series in the parameter
$\epsilon$ :
 $\psi=\psi_0 +\epsilon \psi_1 + \epsilon^2 \psi_2 + \cdots$,
${\partial p}/{\partial x}=({\partial p}/{\partial
 x})_0+\epsilon ({\partial p}/{\partial x})_1 +\epsilon^2 ({\partial
 p}/{\partial x})_2 +\cdots$,
with $u=\partial \psi/\partial y$, $v=-\partial \psi/\partial x$.
The two-dimensional ($x$- and $y$-) momentum equations and the
equation of continuity could be in terms of the stream function
$\psi$ if the $p$-term (the specific body force density, assumed
to be conservative and hence expressed as the gradient of a
time-independent potential energy function) is eliminated. The
final governing equation is
\begin{equation}
 \frac{\partial}{\partial t} \nabla^2 \psi + \psi_y \nabla^2 \psi_x
 -\psi_x \nabla^2 \psi_y =\frac{1}{Re}\nabla^4 \psi,
\hspace*{12mm} \nabla^2 \equiv\frac{\partial^2}{\partial x^2}
+\frac{\partial^2}{\partial y^2}   ,
\end{equation}
and subscripts indicate the partial differentiation. Thus, we have
\begin{equation}
 \frac{\partial}{\partial t} \nabla^2 \psi_0 +\psi_{0y} \nabla^2
 \psi_{0x}-\psi_{0x} \nabla^2 \psi_{0y}=\frac{1}{Re} \nabla^4 \psi_0
 ,
\end{equation}
\begin{equation}
 \frac{\partial}{\partial t} \nabla^2 \psi_1 +\psi_{0y} \nabla^2
 \psi_{1x}+\psi_{1y}\nabla^2 \psi_{0x}-\psi_{0x} \nabla^2 \psi_{1y}-
 \psi_{1x} \nabla^2 \psi_{0y}=\frac{1}{Re} \nabla^4 \psi_1
 ,
\end{equation}
\begin{displaymath}
 \frac{\partial}{\partial t} \nabla^2 \psi_2 +\psi_{0y} \nabla^2
 \psi_{2x}+\psi_{1y}\nabla^2 \psi_{1x}+\psi_{2y} \nabla^2 \psi_{0x}-
\end{displaymath}
\begin{equation}
 \hspace*{25mm} \psi_{0x} \nabla^2 \psi_{2y}- \psi_{1x} \nabla^2 \psi_{1y}-
 \psi_{2x}\nabla^2 \psi_{0y}=\frac{1}{Re} \nabla^4 \psi_2
 ,
\end{equation}
and other higher order terms. The (matter) gas is subjected to
boundary conditions imposed by the symmetric motion of the
vacuum-matter interfaces and the non-zero slip velocity : $u=\mp$
Kn $\,du/dy$ (cf. Refs. 11,12, and 13), $v=\pm
\partial \eta/\partial t$ at $y=\pm (1+ \eta)$, here Kn=mfp$/(2h)$.
The boundary conditions may be expanded in powers of $\eta$ and
then $\epsilon$ :
\begin{displaymath}
 \psi_{0y}|_1 +\epsilon [\cos \alpha (x-t) \psi_{0yy}|_1 +\psi_{1y}|_1
 ]+\epsilon^2 [\frac{\psi_{0yyy}|_1}{2} \cos^2 \alpha (x-t)
 +\psi_{2y}|_1 +
\end{displaymath}
\begin{displaymath}
 \hspace*{3mm}\cos \alpha (x-t) \psi_{1yy}|_1 ] +\cdots =-\mbox{Kn} \{\psi_{0yy}|_1
 +\epsilon [\cos \alpha (x-t) \psi_{0yyy}|_1 +\psi_{1yy}|_1 ]+
\end{displaymath}
\begin{equation}
 \hspace*{3mm}\epsilon^2 [\frac{\psi_{0yyyy}|_1}{2} \cos^2 \alpha (x-t)+
 \cos \alpha (x-t) \psi_{1yyy}|_1 +\psi_{2yy}|_1 ] +\cdots \}
 ,
\end{equation}
\begin{displaymath}
 \psi_{0x}|_1 +\epsilon [\cos \alpha (x-t) \psi_{0xy}|_1 +\psi_{1x}|_1
 ]+\epsilon^2 [\frac{\psi_{0xyy}|_1}{2} \cos^2 \alpha (x-t)+
\end{displaymath}
\begin{equation}
 \hspace*{3mm} \cos \alpha (x-t) \psi_{1xy}|_1 + \psi_{2x}|_1 ] +\cdots =
  -\epsilon \alpha \sin \alpha (x-t).  
\end{equation}
Equations above, together with the condition of symmetry and a
uniform  $(\partial
p/\partial x)_0$, yield :
\begin{equation}
 \psi_0 =K_0 [ (1+2 \mbox{Kn}) y-\frac{y^3}{3}],  \hspace*{24mm}
 K_0=\frac{Re}{2}(-\frac{\partial p}{\partial x})_0 , 
\end{equation}
\begin{equation}
 \psi_1 =\frac{1}{2}\{\phi(y) e^{i \alpha (x-t)}+\phi^* (y) e^{-i \alpha
 (x-t)} \} , 
\end{equation}
where the asterisk denotes the complex conjugate. A substitution
of $\psi_1$ into Eq. (6) yields
\begin{displaymath}
 \{\frac{d^2}{d y^2} -\alpha^2 +i \alpha Re [1-K_0 (1-y^2+2 \mbox{Kn})]\}
 (\frac{d^2}{d y^2} -\alpha^2) \phi -2 i\alpha K_0 Re \,\phi =0  
\end{displaymath}
or
if originally the (matter) gas is quiescent : $K_0 = 0$ (this
corresponds to a free (vacuum) pumping case)
\begin{equation}
 (\frac{d^2}{d y^2} -\alpha^2) (\frac{d^2}{d y^2} -\bar{\alpha}^2) \phi
 =0 ,   \hspace*{24mm} \bar{\alpha}^2= \alpha^2 -i \alpha Re. 
\end{equation}
The boundary conditions are
\begin{equation}
 \phi_y (\pm 1) \pm\phi_{yy} (\pm 1) \mbox{Kn}=2 K_0 (1\pm \mbox{Kn})=0, \hspace*{24mm}
 \phi (\pm 1)=\pm 1 .  
\end{equation}
Similarly, with
\begin{equation}
  \psi_2=\frac{1}{2} \{D(y)+E(y) e^{i 2\alpha (x-t)} +E^* (y) e^{-i
  2\alpha (x-t)} \} ,  
\end{equation}
we have
\begin{equation}
 D_{yyyy}=-\frac{i \alpha Re}{2} (\phi \phi^*_{yy}-\phi^*
 \phi_{yy})_y , 
\end{equation}
\begin{displaymath}
 [\frac{d^2}{d y^2} -(4\alpha^2 -2 i \alpha Re) ] (\frac{d^2}{d y^2} -4\alpha^2)
  E-i 2 \alpha Re K_0 (1-y^2+2 \mbox{Kn})
\end{displaymath}
\begin{equation}
 \hspace*{12mm}(\frac{d^2}{d y^2} -4\alpha^2) E + i 4 \alpha K_0 Re E +
 \frac{i \alpha Re}{2}(\phi_y \phi_{yy}-\phi \phi_{yyy}) =0 ;  
\end{equation}
and the boundary conditions
\begin{equation}
 D_y (\pm 1) +\frac{1}{2} [\phi_{yy} (\pm 1)+\phi^*_{yy} (\pm 1)] -2
 K_0= \mp \mbox{Kn}  \{\frac{1}{2} [\phi_{yyy} (\pm 1) +
\phi^*_{yyy} (\pm 1)]+ D_{yy} (\pm 1) \},  
\end{equation}
\begin{equation}
 E_y (\pm 1)+\frac{1}{2} \phi_{yy} (\pm 1) -\frac{K_0}{2} =\mp \mbox{Kn} [\frac{1}{2}
 \phi_{yyy} (\pm 1) + E_{yy} (\pm 1) ] ,
\end{equation}
\begin{equation}
 \hspace*{24mm} E(\pm 1)+\frac{1}{4} \phi_y  (\pm 1) =0   
\end{equation}
where $K_0$ is zero in Eqns. (16-18).
After lengthy algebraic manipulations, we obtain
 $\phi=c_0 e^{\alpha y}+c_1 e^{-\alpha y}+c_2 e^{\bar{\alpha} y}+
      c_3 e^{-\bar{\alpha} y}$,
where $c_0=(A+A_0)/Det$, $c_1=-(B+B_0)/Det$, $c_2=(C+C_0)/Det$,
$c_3=-(T+T_0)/Det$;
 $Det=A e^{\alpha}-B e^{-\alpha}+C e^{\bar{\alpha}}-T e^{-\bar{\alpha}}$,
\begin{displaymath}
 A=e^{\alpha} \bar{\alpha}^2 (r^2 e^{-2 \bar{\alpha}}-s^2 e^{2
   \bar{\alpha}})-2\alpha \bar{\alpha} e^{-\alpha} w+\alpha \bar{\alpha}
   e^{\alpha} z ( e^{-2\bar{\alpha}} r+ e^{2\bar{\alpha}} s),
\end{displaymath}
\begin{displaymath}
 A_0=e^{-\alpha} \bar{\alpha}^2 (r^2 e^{-2 \bar{\alpha}}-s^2 e^{2
   \bar{\alpha}})+2\alpha \bar{\alpha} e^{\alpha} z-\alpha \bar{\alpha}
   e^{-\alpha} w ( e^{2\bar{\alpha}} s+ e^{-2\bar{\alpha}} r),
\end{displaymath}
\begin{displaymath}
 B=e^{-\alpha} \bar{\alpha}^2 (r^2 e^{-2 \bar{\alpha}}-s^2 e^{2
   \bar{\alpha}})+2\alpha \bar{\alpha} e^{\alpha} z-\alpha \bar{\alpha}
   e^{-\alpha} w ( e^{-2\bar{\alpha}} r+ e^{2\bar{\alpha}} s),
\end{displaymath}
\begin{displaymath}
 B_0=e^{\alpha} \bar{\alpha}^2 (r^2 e^{-2 \bar{\alpha}}-s^2 e^{2
   \bar{\alpha}})-2\alpha \bar{\alpha} e^{-\alpha} w+\alpha \bar{\alpha}
   e^{\alpha} z ( e^{-2\bar{\alpha}} r+ e^{2\bar{\alpha}} s),
\end{displaymath}
\begin{displaymath}
 C=e^{-\alpha} \alpha \bar{\alpha} (w s e^{\bar{\alpha}-\alpha}-r z e^{
   \alpha-\bar{\alpha}})-\alpha e^{2\alpha+\bar{\alpha}} z(\alpha
   z-\bar{\alpha} s)+\alpha e^{-\alpha} w (\alpha
   e^{\bar{\alpha}-\alpha} w-
\bar{\alpha} e^{\alpha-\bar{\alpha}} r),
\end{displaymath}
\begin{displaymath}
 C_0=e^{\alpha} \alpha \bar{\alpha} (w s e^{\bar{\alpha}-\alpha}-r z e^{
   \alpha-\bar{\alpha}})-\alpha z (z\alpha e^{2\alpha-\bar{\alpha}}
   -\bar{\alpha} e^{\bar{\alpha}} s)+\alpha w (\alpha
   e^{-(\bar{\alpha}+2\alpha)} w-
\bar{\alpha} e^{-(2\alpha+\bar{\alpha})} r),
\end{displaymath}
\begin{displaymath}
 T=e^{-\alpha} \alpha \bar{\alpha} (z s e^{\bar{\alpha}+\alpha}-r w e^{-(
   \alpha+\bar{\alpha})})-\alpha \bar{\alpha} (e^{2\alpha-\bar{\alpha}}
   z r -e^{\bar{\alpha}} w s)+
\alpha^2 e^{-\alpha} (-e^{2\alpha} z^2+ e^{-2\alpha} w^2),
\end{displaymath}
\begin{displaymath}
 T_0=e^{\alpha} \alpha \bar{\alpha} (z s e^{\bar{\alpha}+\alpha}-r w e^{-(
   \alpha+\bar{\alpha})})-\alpha \bar{\alpha} (e^{-\bar{\alpha}}
   z r -e^{\bar{\alpha}-2\alpha} w s)+
\alpha^2 e^{\alpha} (-e^{2\alpha} z^2+ e^{-2\alpha} w^2),
\end{displaymath}
with $r= (1-\bar{\alpha} \mbox{Kn})$, $s=(1+\bar{\alpha} \mbox{Kn})$,
$w=(1-\alpha \mbox{Kn})$, $z=(1+\alpha \mbox{Kn})$. \newline To obtain a simple
solution which relates to the mean transport so long as only terms of
$O(\epsilon^2)$ are concerned, we see that if every term in the
x-momentum equation is averaged over an interval of time equal to
the period of oscillation, we obtain for our
solution as given by above equations the time-averaged (unit) body forcing
\begin{equation}
 \overline{\frac{\partial p}{\partial x}}=\epsilon^2 \overline{(\frac{\partial
 p}{\partial x})_2} =\epsilon^2 [\frac{D_{yyy}}{2 Re} + \frac{i Re}{4}
 (\phi \phi^*_{yy} -\phi^* \phi_{yy})] +O (\epsilon^3) =\epsilon^2
 \frac{\Pi_0}{Re} +O(\epsilon^3) ,   
\end{equation}
where $\Pi_0$ is the integration constant for the integration of
equation (15) and could be fixed indirectly in the coming equation
below. Now, from Eq. (17), we have
\begin{equation}
  D_y (\pm 1) \pm \mbox{Kn} D_{yy} (\pm 1)= -\frac{1}{2} [\phi_{yy} (\pm 1)+
  \phi^*_{yy} (\pm 1)] \mp \mbox{Kn}  \{\frac{1}{2} [\phi_{yyy} (\pm 1)
+\phi^*_{yyy} (\pm 1)] \} ,  
\end{equation}
where $D_y (y)= \Pi_0 y^2 +a_1 y+ a_2 + {\cal C} (y)$, and
together from equation (15), we obtain
\begin{displaymath}
 {\cal C} (y)=\frac{\alpha^2 Re^2}{2} [\frac{c_0 c_2^*}{g_1^2}
 e^{(\alpha+\bar{\alpha}^*) y} + \frac{c_0^* c_2}{g_2^2} e^{(\alpha+\bar{\alpha}) y} +
 \frac{c_0 c_3^*}{g_3^2} e^{(\alpha-\bar{\alpha}^*) y} + \frac{c_0^*
 c_3}{g_4^2} e^{(\alpha-\bar{\alpha}) y} +
\end{displaymath}
\begin{displaymath}
 \hspace*{3mm} \frac{c_1 c_2^*}{g_3^2} e^{ (\bar{\alpha}^*-\alpha) y} +
 \frac{c_1^* c_2}{g_4^2} e^{(\bar{\alpha}-\alpha) y}
 + \frac{c_1 c_3^*}{g_1^2} e^{-(\bar{\alpha}^*+\alpha) y}+ \frac{c_1^*
 c_3}{g_2^2} e^{-(\bar{\alpha}+\alpha) y} +
\end{displaymath}
\begin{equation}
 \hspace*{3mm} \frac{c_2 c_3^*}{g_5^2} e^{(\bar{\alpha}-\bar{\alpha}^*) y}+
 \frac{c_2^* c_3}{g_5^2} e^{(\bar{\alpha}^* -\bar{\alpha}) y} +
 2 \frac{c_2 c_2^*}{g_6^2}  e^{(\bar{\alpha}^*
 +\bar{\alpha}) y}+2 \frac{c_3 c_3^*}{g_6^2}  e^{-(\bar{\alpha}^*
 +\bar{\alpha}) y}]  ,
\end{equation}
with $g_1=\alpha+\bar{\alpha}^*$, $g_2=\alpha+\bar{\alpha}$,
$g_3=\alpha-\bar{\alpha}^*$, $g_4=\alpha-\bar{\alpha}$,
$g_5=\bar{\alpha}-\bar{\alpha}^*$,
$g_6=\bar{\alpha}+\bar{\alpha}^*$. In realistic applications we
must determine $\Pi_0$ from considerations of conditions at the ends
of the matter-region. $a_1$ equals to zero because of the symmetry of
boundary conditions.  \newline Once $\Pi_0$ is specified,
our solution for the mean speed ($u$ averaged over time) of matter-flow
is
\begin{equation}
 {U}=\epsilon^2 \frac{D_y}{2}=\frac{\epsilon^2}{2} \{ {\cal
 C}(y)-{\cal C} (1)+R_0-\mbox{Kn} \,{\cal C}_y (1)+\Pi_0 [y^2-(1+2 \mbox{Kn})] \} 
\end{equation}
where $R_0$ $=-\{ [\phi_{yy} (1)+\phi^*_{yy} (1)]$ $- \mbox{Kn}
[\phi_{yyy} (1)$ $+\phi^*_{yyy} (1)]\}/2$, which has a numerical
value about $3$ for a wide range of $\alpha$ and $Re$ (playing the role of viscous
dissipations) when Kn$=0$. To illustrate our results clearly, we adopt $U(Y) \equiv u(y$)
for the time-averaged results with $y\equiv Y$ in the following.
\section{Results and Discussion}
We check our approach firstly by examining $R_0$ with that of
no-slip (Kn$=0$) approach. This can be done easily once
we consider terms of $D_y(y)$ and ${\cal C} (y)$ because to
evaluate $R_0$ we shall at most take into account the higher
derivatives of $\phi(y)$, like $\phi_{yy} (y)$, $\phi_{yyy} (y)$
instead of $\phi_y (y)$ and escape from the prescribing of $a_2$.
\newline Our numerical calculations confirm that the mean streamwise
velocity distribution (averaged over time) due to the induced
motion by the wavy elastic vacuum-matter interface in the case of
free (vacuum) pumping is dominated by $R_0$ (or Kn) and the
parabolic distribution $-\Pi_0 (1-y^2)$. $R_0$ which defines the
boundary value of $D_y$ has its origin in the y-gradient of the
first-order streamwise velocity distribution, as can be seen in
Eq. (17).
\newline In addition to the terms mentioned above, there is a
perturbation term which varies across the channel : ${\cal C} (y)
-{\cal C} (1)$. Let us define it to be
\begin{equation}
 F(y)=\frac{-200}{\alpha^2 Re^2} [{\cal C} (y) -{\cal C} (1)]
\end{equation}
 We remind the readers that the Reynolds number
here is based on the wave speed.
The physical trend herein is also the same as those reported in
Refs. 12 and 13 for the slip-flow effects. The slip produces
decoupling with the inertia of the wavy interface.
\newline Now, let us define a critical reflux condition as one for
which the mean velocity ${U} (Y)$ equals to zero at the
center-line $Y=0$ (cf. Fig. 2). With equations (15,23-24), we have
\begin{equation}
  \Pi_{0_{cr}}=Re \overline{(\frac{\partial p}{\partial x})_2}=\frac{[{\alpha^2
  Re^2}F(0)/200 +\mbox{Kn} \,{\cal C}' (1)-R_0]}{-(1+2 \mbox{Kn})}
\end{equation}
which means the critical reflux condition is reached when $\Pi_0$
has above value. Pumping against a positive (unit) body forcing
greater than the critical value would result in a backward
transport (reflux) in the central region of the stream. This
critical value depends on $\alpha$, $Re$, and Kn. There will be no
reflux if the (unit) body forcing or pressure gradient is smaller
than this $\Pi_0$. Thus, for some $\Pi_0$ values less than
$\Pi_{0_{cr}}$, the matter (flow) will keep moving or evolving
forward. On the contrary, parts of the matter (flow) will move or
evolve backward if $\Pi_0 > \Pi_{0_{cr}}$. This result could be
similar to that in Ref. 17 using different approach or
qualitatively related to that of Ref. 4 : even for very slow
growth of $\Lambda$, the gravitationally bound systems become
unbound while the nongravitationally bound systems remain bound
for certain parameters defined in Ref. 4 (e.g., $\eta$). \newline
We present some of the values of $\Pi_0(\alpha, Re; \mbox{Kn}=0,
0.1)$ corresponding to dark  states which satisfy $\int_{-1}^1
U(Y) dY=0$ in  Table 1 where the wave number ($\alpha$) has the
range between $0.20$ and $0.80$; the Reynolds number
($Re$)$=0.1,1,10,50,100$.  Here, the dark state is directly
related to the zero-mass-flux (in average) state in bulk sense
although there are locally velocity differences across the
(confined) space. It means the gravity effects are not absent
locally considering the velocity distribution across the confined
space. To be specific, the dark state is difficult to be observed
since there is  no mass flux even though this state happens!
\newline We observe that, from Table 1, as Kn increases from zero
to 0.1, the critical $\Pi_0$ or time-averaged (unit) body forcing
decreases significantly. For the same Kn, once Re is larger than
10, critical reflux values $\Pi_0$ drop rapidly and the
wave-modulation effect (due to $\alpha$) appears. The latter
observation might be interpreted as the strong coupling between
the vacuum-matter boundary and the inertia of the streaming
matter-flow. The illustration of the velocity fields for those
dark states are shown in Figure 2. There are three wave numbers :
$\alpha=0.2, 0.5, 0.8$. The Reynolds number (Re) is $50$. Both
no-slip and slip (Kn=0.1) cases are presented in Fig. 2. The line
of value $U\equiv0$ in Fig. 2 is schematic and could represent the
direction of positive and negative velocity fields.
\newline Some remarks could be made about these dark states (or solitons) : the
matter or universe being freezed in the time-averaged sense for
specific dissipations (in terms of Reynolds number which is the
ratio of wave-inertia and viscous effects) and wave numbers (due
to the wavy vacuum-interface or vacuum fluctuations) for either
no-slip and slip cases. This particular result might also be
related to a changing cosmological term (growing or decaying
slowly) or the critical density mentioned in Ref. 2. If we treat
the (unit) body forcing as the pressure gradient, then for the
same transport direction (say, positive x-direction), the negative
pressure (either downsdtream or upstream) will, at least, occur
once the time-averaged flow (the maximum speed of the matter (gas)
appears at the center-line) is moving forward!
\newline Meanwhile, the time-averaged transport induced by the wavy interface
is proportional to the square of the amplitude ratio (although the
small amplitude waves being presumed), as can be seen in Eqn. (12)
or (20), which is qualitatively the same as that presented in Ref.
9 for analogous interfacial problems. In brief summary, the
entrained transport (pattern, either postive or negative and there
being possible dark states) due to the wavy vacuum-matter boundary
 is mainly tuned
by the (unit) body forcing or $\Pi_0$ for fixed Re (viscous
dissipation). Meanwhile, $\Pi_{0_{cr}}$ depends strongly on the
Knudsen number (Kn, a rarefaction measure) instead of Re or
$\alpha$ (wave number). We hope that in the future we can
investigate other issues [22-26]
 using the present or more advanced approach.

\begin{table}[h]
 \caption{Dark states values ($\Pi_0$) for a flat vacuum-matter boundary.}
\vspace*{5mm}
\begin{center}
\begin{tabular}[b]{|r|c|c|c|c|c|c|}      \hline
    &          &  Re  &   &     &     &      \\ \hline
 Kn & $\alpha$ &  0.1 & 1 & 10 & 50 &  100 \\ \hline
 0 
    & 0.2  &  4.5269  & 4.5269  &   4.5231       &   4.4496   &   4.3275  \\  \cline{2-7}
    & 0.5  &  4.6586  & 4.6584  &   4.6359       &  4.4086    &  4.2682  \\  \cline{2-7}
    & 0.8  &   4.9238 &  4.9234 &    4.8708      &  4.5714   &   4.4488  \\ \hline 
0.1 
     & 0.2 &  2.4003  & 2.4000  &   2.3774       &  1.9532    &   1.2217    \\ \cline{2-7}
     & 0.5 &  2.4149  & 2.4132  &   2.2731       &  0.7728    &    -0.9054 \\
     \cline{2-7}
     & 0.8 &  2.4422  & 2.4379  &   2.0718       &  -0.5885   &    -3.4151  \\ \hline 
\end{tabular}            
\end{center}
\end{table}


\psfig{file=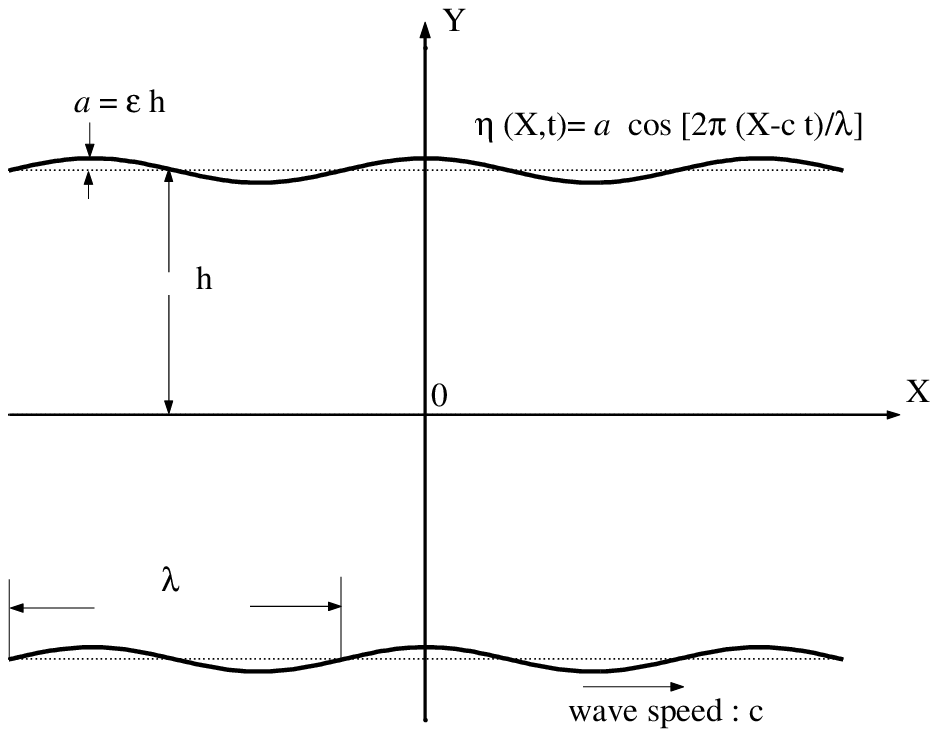,bbllx=0.0cm,bblly=14.8cm,bburx=12cm,bbury=24cm,rheight=9.2cm,rwidth=9.2cm,clip=}

\begin{figure} [h]
\hspace*{3mm} Fig. 1 \hspace*{1mm} Schematic diagram of the
deformable motion of the vacua-matter boundary.
\end{figure}

%

%
%
%
\vspace{3mm}
\psfig{file=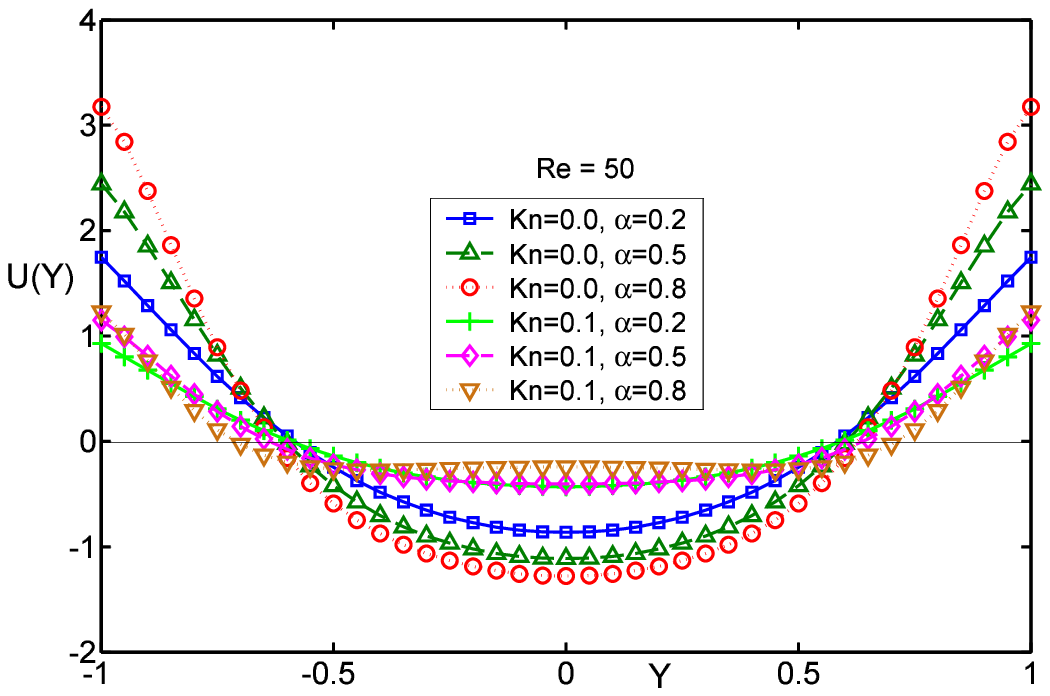,bbllx=0.0cm,bblly=19.5cm,bburx=12cm,bbury=28cm,rheight=8cm,rwidth=8cm,clip=}
%
\begin{figure} [h]
\hspace*{3mm} Fig. 2 \hspace*{1mm} Demonstration of the dark
states : the mean velocity field $U(Y)$ for \newline \hspace*{3mm}
wave numbers $\alpha=0.2,0.5,0.8$. The Reynolds number is $50$. Kn
is the rarefaction measure
\newline \hspace*{3mm}
(the mean free path of the particles divided by the characteristic length).
\newline \hspace*{3mm}
The $U\equiv0$ line is schematic and illustrates the directions of
positive and negative $U(Y)$.
\newline \hspace*{3mm}
The integration of $U(Y)$ w.r.t. $Y$ for these velocity fields
gives zero volume (mass) flow rate.
\end{figure}

\end{document}